\documentclass[reprint,amsmath,amssymb,aps,showpacs,prl,longbibliography]{revtex4-1}
\usepackage{ctable}
\usepackage{slashbox}

\usepackage{graphicx}
\usepackage{dcolumn}% Align table columns on decimal point
\usepackage{bm}% bold math
\usepackage[colorlinks=true]{hyperref}% add hypertext capabilities
\usepackage{mathrsfs}
\usepackage{amsthm}
\usepackage{bm}

\newcommand{\vect}[1]{\mathbf{#1}}

\newcommand{\ket}[1]{|#1\rangle}
\newcommand{\bra}[1]{\langle #1|}

\renewcommand{\bf}[1]{\textbf{#1}}
\def\be{\begin{equation}}
\def\ee{\end{equation}}
\def\bea{\begin{eqnarray}}
\def\eea{\end{eqnarray}}

\usepackage{xcolor}

\begin{document}

\title{Topologically protected quantization of work}

\author{Bruno Mera}
\email{bruno.mera@tecnico.ulisboa.pt}
\affiliation{Instituto de Telecomunica\c{c}\~{o}es, Lisboa}
\affiliation{Instituto Superior T\'{e}cnico, Universidade de Lisboa, Portugal}

\author{Krzysztof Sacha}
\affiliation{Instytut Fizyki imienia Mariana Smoluchowskiego, 
Uniwersytet Jagiello\'nski, ulica Profesora Stanis\l{}awa \L{}ojasiewicza 11, PL-30-348 Krak\'ow, Poland}

\author{Yasser Omar}
\affiliation{Instituto de Telecomunica\c{c}\~{o}es, Lisboa}
\affiliation{Instituto Superior T\'{e}cnico, Universidade de Lisboa, Portugal}

\date{\today}% It is always \today, today,

\begin{abstract}
The transport of a particle in the presence of a potential that changes periodically in space and in time can be characterized by the amount of work needed to shift a particle by a single spatial period of the potential. In general, this amount of work, when averaged over a single temporal period of the potential, can take any value in a continuous fashion. Here we present a topological effect inducing the quantization of the average work. We find that this work is equal to the first Chern number calculated in a unit cell of a space-time lattice. Hence, this quantization of the average work is topologically protected. We illustrate this phenomenon with the example of an atom whose center of mass motion is coupled to its internal degrees of freedom by electromagnetic waves.
\end{abstract}

\maketitle

%\tableofcontents

%\section{Introduction}
%\label{sec:Introduction}

Topological phases of matter constitute a new paradigm in condensed matter physics. Remarkable examples are the Haldane anomalous insulator~\cite{hal:88}, an  instance of the more general Chern insulators, and, even more generally, topological insulators and superconductors~\cite{ber:hug:13,has:kan:10}. The later are symmetry protected topological phases of free fermions. Unlike the conventional phases of matter described by the Landau-Ginzburg theory in terms of a local order parameter~\cite{and:97}, instead, topological insulators and superconductors are described by topological invariants, such as the holonomy of a flat connection, like the Zak phase~\cite{zak:89, ber:84}, or the Chern number of a vector bundle~\cite{nak:03,mor:01,ste:99,bot:tu:13} over a $2$-torus or an arbitrary Riemann surface, like the Thouless-Kohmoto-Nightingale-den Nijs (TKNN) invariant~\cite{tknn:82}. These topological invariants measure the non-trivial ``twisting'' of the wave-functions of the bulk, which are usually subject to certain generic symmetries like time-reversal, particle-hole or chiral symmetries. Topological insulators and superconductors  were systematically classified, using \emph{K}-theory~\cite{kar:08}, by Kitaev~\cite{kit:09}; and using homotopy groups and Anderson localization, by Schnyder, Ryu, Furusaki and Ludwig~\cite{sch:ryu:fur:lud:08, sch:ryu:fur:lud:09}. The resulting classification exhibits Bott-periodicity, $2$-fold for the complex case and $8$-fold for the real case, and is known as the periodic table of topological insulators and superconductors. Moreover, the bulk-to-boundary principle predicts that, when terminating the system to the vacuum, there will appear gapless modes living in the boundary of the system. These modes are topologically protected. One can understand the existence of these gapless modes by anomaly inflow arguments~\cite{ryu:moo:lud:12, wit:16}.

Topological insulators and superconductors~\cite{has:kan:10, qi:zha:11, ber:hug:13} are very attractive from the experimental point of view due to their robustness to perturbations and also due to the many potential applications to photonics, spintronics, quantum computing, and, more generally, to the emergent field of quantum technologies~\cite{xu:mio:liu:14,orn:16}.

Experimentally, one can study topological insulators and superconductors in quantum simulators which are versatile systems that can mimic behavior of other systems difficult to control in the laboratory. Among the physical platforms for the quantum simulation of topological matter, ultracold atoms in optical lattices~\cite{gol:juz:ohb:spie:14,gol:bud:zol:16} and topological photonics~\cite{oza:pri:amo:gol:haf:lu:rec:sch:sim:zil:18} offer the most promising realizations. Quantum simulators have allowed for the realization of the topological insulators in one-dimensional (1D)~\cite{car:der:dau:17,kit:bro:fed:12,mei:an:gad:16,ata:aid:bar:aba:kit:dem:blo:13,stj:gob:gal:lem:oza:leg:sag:blo:amo:17,mei:an:dau:maf:mas:hug:gad:18}, 2D~\cite{aid:loh:sch:ata:bar:nas:coo:blo:gol:15,stu:lu:ayc:gen:spi:15,tar:una:fla:rem:eck:sen:wei:17} and even 4D space~\cite{loh:sch.pri:zil:blo:18,zil:hua:gug:wan:che:kra:rec:18} -- the latter with the help of a synthetic dimension.

Thirty years ago Thouless proposed the idea of a topological charge pump where transport of charge, described by an adiabatically and periodically evolving Hamiltonian, was quantized and determined by the first Chern number calculated in the time-momentum space~\cite{thou:83}. More concretely, if one has a one-dimensional translation invariant gapped system of free fermions on a lattice then, by adiabatically and periodically driving the system, the center of mass position is shifted, in one period of driving, by an integer multiple of the lattice constant. This integer is the first Chern number of the vector bundle of occupied states in the instantaneous ground states of the system, defined over the space of Bloch momenta and time --  topologically, a torus. Direct observation of the Thouless quantum pump was demonstrated in a quantum simulator where bosonic ultra-cold atoms were prepared in the Mott insulator phase in an optical lattice whose tunneling amplitudes were periodically modulated in time~\cite{loh:sch:chr:zil:aid:bloc:16,Shuta2016}.

In the present letter we consider a different phenomenon. Namely, we consider an atom constrained to move in $1$D, with internal degrees of freedom subject to a space-time periodic potential coupling the internal states. In this case, it is possible, by preparing the system in a dressed state, that the atom experiences an effective synthetic electric field whose average work, in a period of driving and one wavelength, is quantized in units of the Planck constant $h$. The quantization is topological in nature and robust against deformations of the system preserving the gap. In the following, we provide an explicit situation where this topological effect occurs and propose a way to experimentally realize it. The differences between this phenomenon and that of Thouless pumping are pointed out in Table~\ref{tab: comparison}.
  
Let us consider an atom where the ground state energy level is characterized by the total angular momentum $F=1$ and in the presence of an external magnetic field the energies $E_{m_F}$, $m_F=+1,0,-1$ of the magnetic sublevels are split, $\Delta E=E_1-E_0=E_0-E_{-1}$. We denote the internal states by $\ket{0}\equiv \ket{m_F=0}$, $\ket{1}\equiv \ket{m_F=+1}$ and $\ket{2}\equiv \ket{m_F=-1}$. If an atom is subjected to two counter-propagating circularly polarized electromagnetic waves of the frequency $\omega$, the internal degrees of freedom of an atom and the electromagnetic fields can be described by the dressed-atom Hamiltonian which, within the rotating wave approximation, reads~\cite{gol:juz:ohb:spie:14},
\bea
M(t,x)&=&\delta(t)\left(\ket{1}\bra{1}-\ket{2}\bra{2}\right)
+\Omega(t,x)\ket{0}\bra{1}\cr &&+\Omega^*(t,x)\ket{0}\bra{2}+{\rm H.c.},
\eea
where we assume that the detuning is oscillating in time due to the periodic modulation of the magnetic field, $\delta(t)=\Delta E(t)-\hbar\omega =\gamma +\nu \cos(\widetilde{\omega} t)$, with the frequency $\tilde\omega\ll\omega$ and $\nu,\gamma\in\mathbb{R}$. The Rabi frequency depends periodically on time and space, $\Omega(t,x)=\alpha_1(t)e^{ikx}+\alpha_2(t)e^{-ikx}$ where $k$ denotes the wave number of the electromagnetic waves while $\alpha_1(t)=(\alpha/2)\cos(\widetilde{\omega}t -\pi/4)$ and $\alpha_2(t)=(\alpha/2)\cos(\widetilde{\omega}t -\pi/4)$ describe periodic modulations of the amplitudes of the waves, with the same frequency as the frequency of the magnetic field modulation, where $\alpha$ is proportional to the dipole matrix element. The Hamiltonian $M(t,x)$ is periodic both in space and in time with the periods $\lambda=2\pi/k$ and $T=2\pi/\tilde\omega$, respectively, and can be written in a more compact form, $M(t,x)=\sum_{\mu=1}^3B^{\mu}(t,x)J_{\mu}$ where $J_3=\ket{1}\bra{1}-\ket{2}\bra{2}$, $J_1-iJ_2=\sqrt{2}(\ket{0}\bra{1}+\ket{2}\bra{0})$ and
\bea
B^1(t,x)&=&\alpha\cos(kx) \cos(\tilde\omega t), \cr
B^2(t,x)&=&\alpha\sin(kx) \sin(\tilde\omega t), \cr
B^3(t,x)&=&\gamma+\nu\cos(\tilde\omega t).
\label{eq: B}
\eea

When the atomic center of mass motion is coupled to its internal degrees of freedom certain geometric gauge fields arise~\cite{Juzeliunas2004,Dalibard2011,rus:juz:ohb:fle:05,gol:juz:ohb:spie:14,dar:jam:12}. For simplicity, let us consider that the atomic motion is restricted to one spatial dimension. The full Hamiltonian of the system is given by
\begin{eqnarray}
H=\frac{p^2}{2m}+M(t,x),
\label{fullH}
\end{eqnarray}
where $x$ and $p$ are the atomic center of mass coordinate and momentum, $m$ is the mass.
We can solve the eigenvalue problem for $M(t,x)$, yielding the eigenvalues $\varepsilon_1(t,x)=|\vect B(t,x)|=\sqrt{\sum_{\mu}(B^\mu)^2}$, $\varepsilon_2(t,x)=0$ and $\varepsilon_3(t,x)=-\varepsilon_1(t,x)$ and the corresponding eigenstates (dressed states of an atom) $\ket{\eta_{i}(t,x)}$. 
%Because $(t,x)\in\mathbb{R}^2$, which is contractible, and $M(t,x)$ is smooth, $\ket{\eta_{i}(t,x)}$'s are globally defined and also smooth. 
The most general solution of the Schr\"{o}dinger equation will be given by a linear combination 
$\psi(t,x)=\sum_{i=1}^{3}\Psi^i(t,x)\ket{\eta_i(t,x)}$.
Writing the vector $\Psi=(\Psi^1,\Psi^2,\Psi^3)^T$, we get the time-dependent Schr\"{o}dinger equation corresponding to the Hamiltonian (\ref{fullH}) in the form
\begin{eqnarray}
\Big[i\hbar\Big(\frac{\partial}{\partial t}+\mathcal{A}_0\Big)+\frac{\hbar^2}{2m}\Big(\frac{\partial}{\partial x}+\mathcal{A}_{1}\Big)^2- V -\mathcal{E}\Big]\Psi=0,
\end{eqnarray}
where $\mathcal{E}=\mbox{diag}(\varepsilon_{1},\varepsilon_2,\varepsilon_{3})$ and the matrices $\mathcal{A}_0=\big(\bra{\eta_i}\partial_t\ket{\eta_j}\big)$ and $\mathcal{A}_1=\big(\bra{\eta_i}\partial_x\ket{\eta_j}\big)$ are the components of the matrix-valued one-form $\mathcal{A}=\big(\bra{\eta_i}d\ket{\eta_j}\big)$.

For the configuration of the electromagnetic waves and the detuning we have chosen, the eigenvalues $\varepsilon_i(t,x)$ are separated from each other by gaps for each $(t,x)$. If we prepare an atom in, e.g., the positive energy dressed state, 
\be
\ket{\eta_1(t,x)}\! = \! \frac{\ket{1} \! +\! \sqrt{2} z \ket{0} \! +\! z^2\ket{2}}{1+|z|^2},\! \ z(t,x)=\frac{B^1+iB^2}{|\bf{B}|+B^3},
\ee
it will follow this state in the time evolution provided its kinetic energy is much smaller than the gap between the adjacent dressed state levels. Then, one can perform an adiabatic Born-Oppenheimer approximation and project the dynamics onto this state only~\cite{rus:juz:ohb:fle:05,Dalibard2011,gol:juz:ohb:spie:14,dar:jam:12}. The resulting effective Schr\"odinger equation for the center of mass wave-function, $\phi(t,x)=\Psi^1(t,x)$, is that of a particle in the presence of an external gauge field, $A_0/\hbar=i\bra{\eta_{1}}\partial_t\ket{\eta_1}$ and $A_1/\hbar=i\bra{\eta_{1}}\partial_x\ket{\eta_1}$, and an effective scalar potential $V_{\rm eff}$,
\begin{eqnarray}
i\hbar\frac{\partial\phi}{\partial t}=\left[\frac{1}{2m}\left(p-A_{1}\right)^2- A_0+ V_{\text{eff}}\right]\phi,
\label{eq:schrod_eff}
\end{eqnarray}
with 
\begin{eqnarray}
V_{\text{eff}}=\frac{\hbar^2}{2m}g_{11}(t,x)+\varepsilon_1(t,x),
\end{eqnarray}
where $g_{11}=\sum_{j>1}|\bra{\eta_1}(\partial M/\partial x)\ket{\eta_j}|^2/(\varepsilon_j-\varepsilon_1)^2$ is the $11$th component of the \emph{quantum metric}~\cite{dar:jam:12,rest:11,kol:sel:meh:pol:17}. 

Since the atom is constrained to move in a single space dimension, the only relevant component of the field strength tensor is the synthetic electric field force acting on a particle of a unit charge
\begin{align}
E(t,x)=\frac{\partial A_1}{\partial t}-\frac{\partial A_0}{\partial x}=\hbar 
\frac{\vect B\cdot\big(\frac{\partial \vect B}{\partial x}\times \frac{\partial \vect B}{\partial t}\big)}{|\vect B|^3}.
\end{align}
Because $\vect B\equiv (B^1,B^2,B^3)$ is space-time periodic, one can define a Chern number $c_1$ associated to the positive energy dressed state which will be minus twice the winding number of the map $(t,x)\mapsto \vect{B}(t,x)/|\vect{B}(t,x)|\in S^2$, where $S^2$ denotes the unit sphere in $\mathbb{R}^3$, i.e. $c_1=\frac{1}{2\pi\hbar}\int_{0}^{T}\int_{0}^{\lambda} E(t,x)dt dx$. In particular, with $\alpha/\nu=1$, using Eq.~\eqref{eq: B}, we get a nontrivial Chern number $-4$ for $-1<\gamma/\nu<0$ and $4$ for $0<\gamma/\nu<1$ and trivial elsewhere \cite{Sup1w}. Similar result holds for the negative energy dressed state but with the Chern number being the opposite. The zero energy dressed state always has trivial Chern number. 

The quantization of the Chern number, proved in \cite{Sup1w}, amounts to having, on the unit cell of the space-time lattice, a quantized value for the flux $\int E(t,x)dt dx $ in units of $2\pi\hbar \equiv h$. Now $E(t,x)dx$ is, dimensionally, the amount of work, of the {\it electric field force}, under the displacement $dx$ of a particle with a unit charge. The space-time lattice involved is simply $\Lambda=\{(t,x)=(m T, n \lambda) ,\ m,n\in\mathbb{Z}\}$. The interpretation of the quantized value of the Chern number is the following: the average over a period $T$ of the work performed by the {\it electric field} $E$ in the transport of a classical particle by a distance of a single space cell, i.e. $x\rightarrow x+\lambda$, is quantized in units of Planck's constant $h$. If we consider the normalized average in time, to have proper units of work, we get $(1/T)\int_{0}^{T}\int_{0}^{\lambda} E(t,x)dt dx= (h/T) c_1=(\hbar \widetilde{\omega}) c_1$, with $c_1\in\mathbb{Z}$ the Chern number. We thus get quantization in units of the driving energy $\hbar \widetilde{\omega}$.

\begin{figure}[h!]
\centering
\includegraphics[scale=0.45]{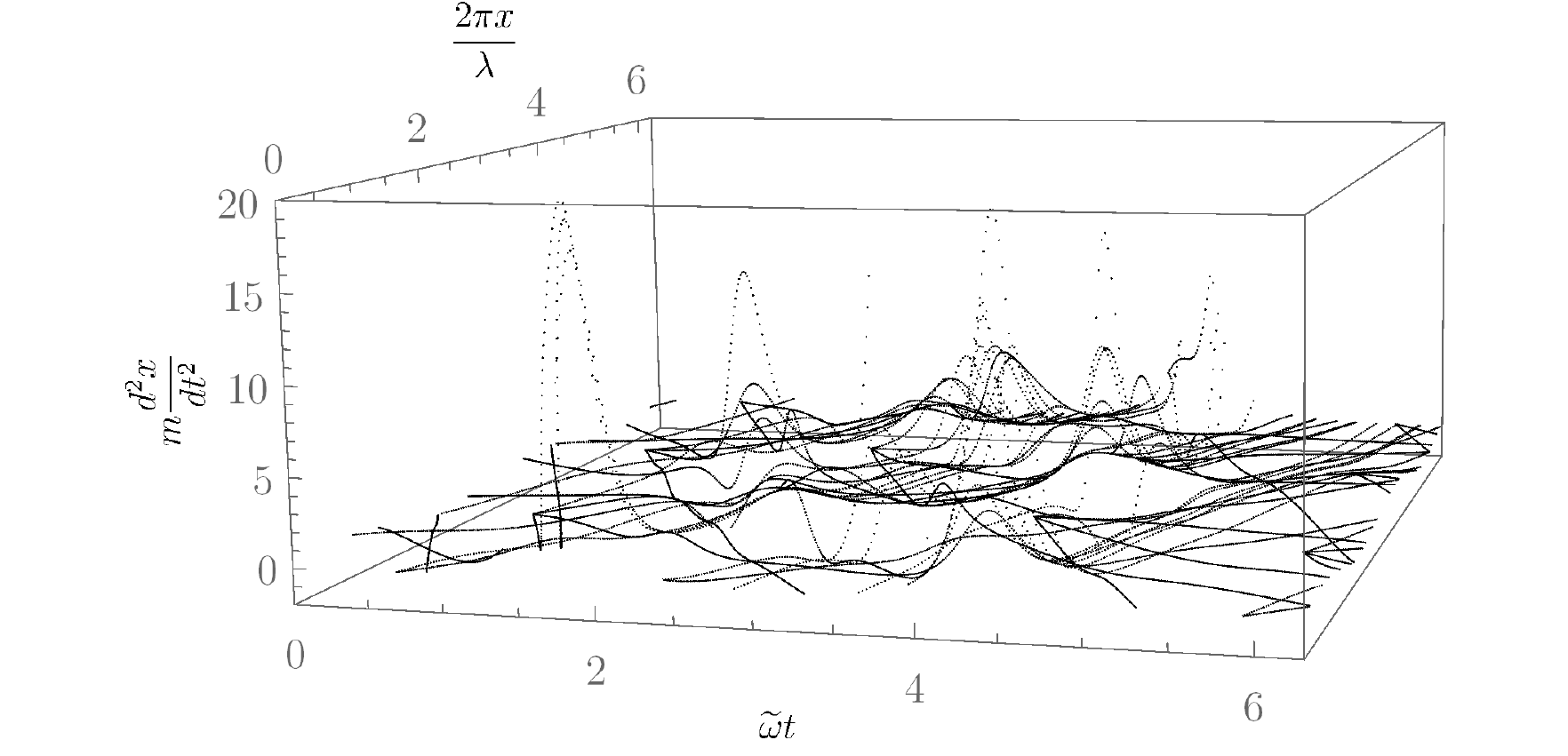}
\includegraphics[scale=0.45]{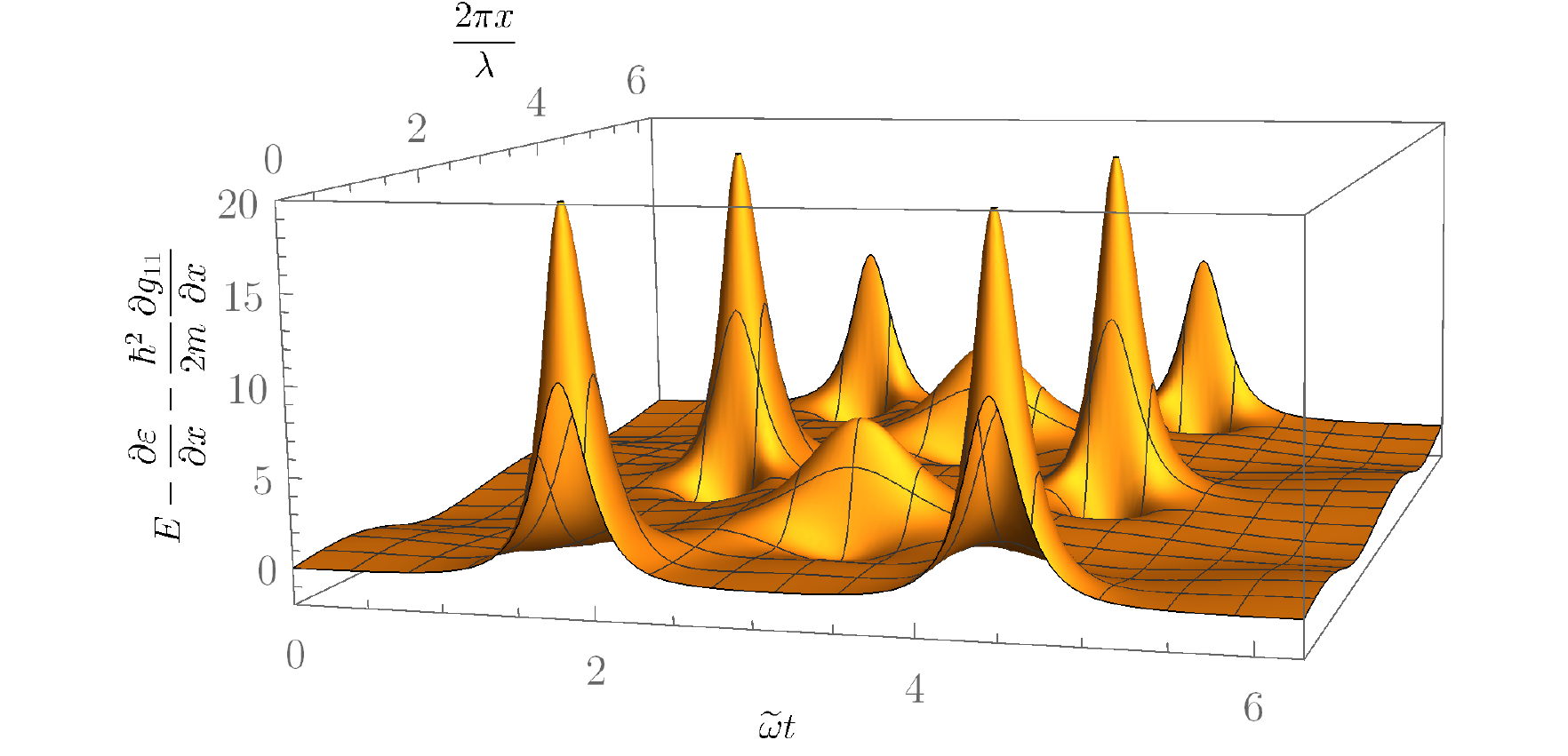}
\includegraphics[scale=0.45]{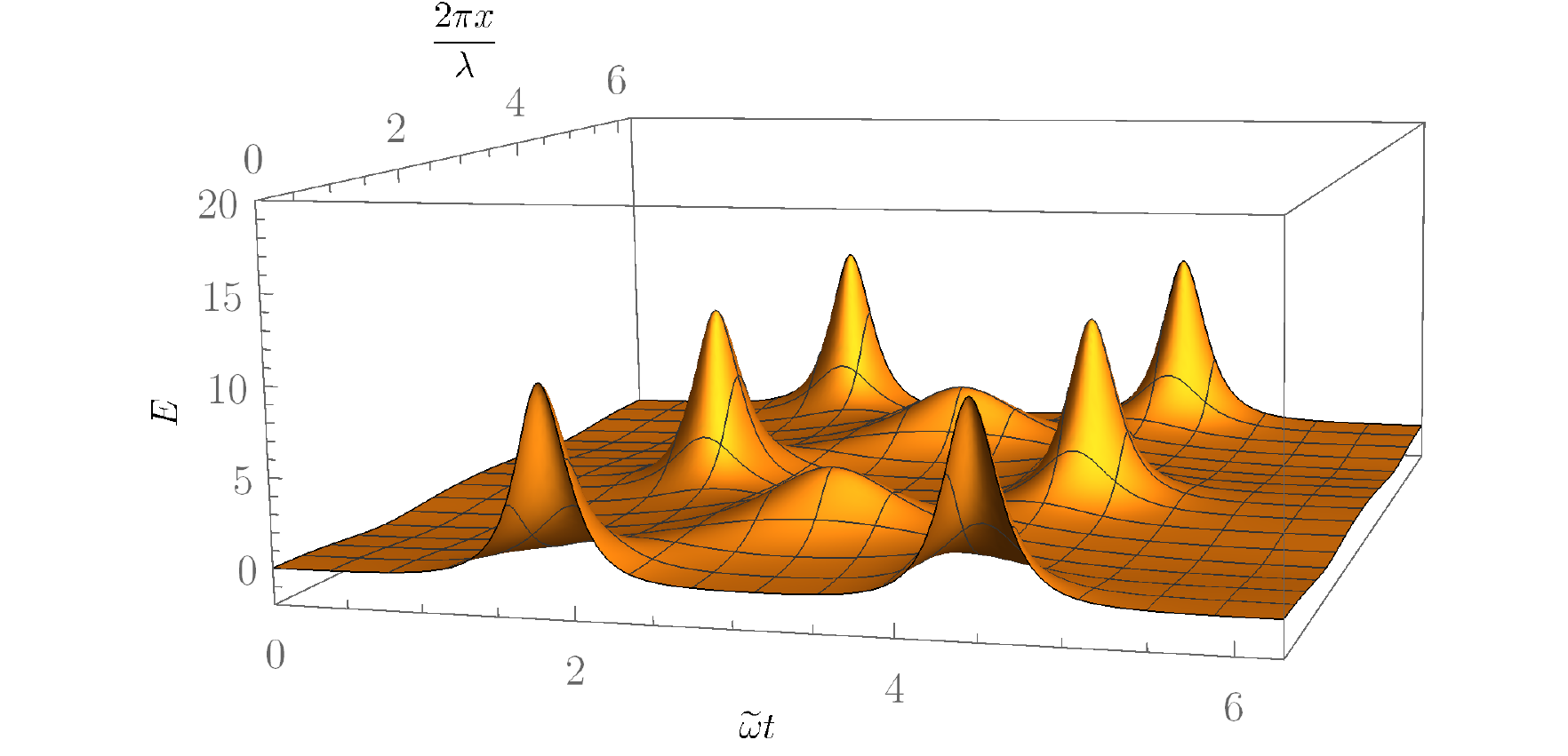}
\caption{Proposal for experimental demonstration of the quantization of the average work performed by the {\it electric field} $E(t,x)$. The top panel shows acceleration profile in a space-time unit cell obtained by integration of the classical equations of motion~\eqref{classeq}. For each initial condition, the resulting classical trajectory allows one to calculate the acceleration by differentiation of the trajectory with respect to time twice.  In the middle panel we show the same profile obtained by plotting the total force acting on an atom. In the bottom panel the contribution from the {\it electric field force} $E(t,x)$ is plotted only. The integration of $E(t,x)$ over a single space-time cell results in a quantized value which can be also obtained when one estimates the integral with the help of the points presented in the top panel. In the latter case, one has to first subtract the other contributions to the total force which are known from the theoretical description, cf. Eq.~\eqref{classeq}. Here $\mu=0.5$ and $m=1$. The Chern number is $4$. Moreover, we take $\hbar=1$.}
\label{fig: acc and force}
\end{figure}

A possible way to experimentally observe quantization of the average work, in the example we consider, can be done indirectly as follows. Take the time interval $[0,T]$ and consider a number $N$ of instants $t_{i}$, $i=1,...,N$. For each instant $t_i$, prepare an atom in the dressed state band with energy $\varepsilon_1(t_i,x)$ and described, at $(t_i,x)$, by $\ket{\eta_1(t_i,x)}$. We want the state of the center of mass degree of freedom of an atom to be strongly localized in a certain point $x(t_i)$ (i.e. much better than the size of a single space cell which is not a problem if the experiment is performed in the RF range where the wavelength $\lambda$ is of the order of the meter), 
so the dynamics for time $t\in[t_i,T]$ is well described by the classical equations of motion:
\begin{eqnarray}
m\frac{d^2 x}{dt^2}(t)&=& E(t,x(t))  \nonumber \\
&&-\frac{\partial \varepsilon_1}{\partial x} (t,x(t)) -\frac{\hbar^2}{2m}\frac{\partial g_{11}}{\partial x} (t,x(t)).
\label{classeq}
\end{eqnarray}
We then measure the position of the atoms in the period $[0,T]$. 
With the resulting trajectories $x_{i}(t)$, $t\in [t_i,T]$, $i=1,...,N$, we can  then differentiate with respect to time twice obtaining the acceleration. With this procedure we will get a profile of the total force field in the unit cell of $\Lambda$ which we can compare to the theoretical predictions. Due to the localization of the center of mass of an atom, the observed profile should be the same and quantization of the time average of work of $E(t,x)$ can be confirmed.
In Fig.~\ref{fig: acc and force} we show how the sampling of the accelerations of trajectories allows us to have the force profile on the unit cell. Additionally, we show the profile of $E(t,x)$ on the unit cell. The total force and the profile of $E(t,x)$ are qualitatively similar. The reason is that all contributions to the total force increase with the decrease of the gap function $\varepsilon_1(t,x)$.

We would like to remark that the phenomenon we describe is at the boundary between classical and quantum physics. Classical, since we want the states of the atoms to be strongly localized, so that the dynamics is classical. This is achieved by staying in the cold atom regime and not in the ultracold one. Quantum, since the atom will experience the effect of a \emph{synthetic force} whose average work on the unit cell is quantized due to the quantum nature of the wave-function of the internal degrees of freedom of the atom. This is achieved by having a gap which is larger than the kinetic energy of the atom.

%\re{Finally, we would like to}
We stress that the topological effect of work quantization considered in this letter and that of Thouless pumping are physically different, although mathematically similar, cf.~Table~\ref{tab: comparison}. Explicitly, in our case, it is the topology of quantum states over space-time and not of quantum states over Bloch momentum space that is involved. This topology is then reflected in the \emph{quantization} of the average work of the synthetic electric field and not the quantization of the shift of the centre of mass of the system. 

Finally, we would like to make contact with the very recent Ref.~\cite{kol:fred:gaz:mor:moo:18} 
%, pointed out by one of the Referees,
where a topological energy pump in $1D$, in the context of a driven system was considered. There, a ``work polarization'' is quantized. We remark that the topological invariant there refers to the homotopy class of a map $\mathcal{P}$, describing the dynamics within each cycle, from a three-dimensional torus to the unitary group $\mbox{U}(2)$. The three-dimensional torus is parametrized by variables $(t,\lambda,k)$ where $t$ is time, $\lambda$ is a flux and $k$ is the one-dimensional momentum in the first Brillouin zone. As a consequence, although in both cases there is quantization of some type of work, just as the Thouless pumping is significantly different from the phenomenon considered here, see Table~\ref{tab: comparison}, so is this one.

In summary, we have presented an effect in which the transport of a particle in the presence of a space-time periodic potential is characterised by a quantized average, over a period of the potential, amount of work needed to shift a particle by a single spatial period of the potential. The quantization was understood in terms of the topological twist of the vector bundle of dressed states. Moreover, we have provided an experimental procedure to probe this phenomenon.\\

We are grateful to Tomasz Kawalec for a fruitful discussion concerning experimental aspects. B.M. and Y.O. thank the support from Funda\c{c}\~{a}o para a Ci\^{e}ncia e a Tecnologia (Portugal), namely through programme POCH and projects UID/EEA/50008/2013, UID/EEA/50008/2019 and  IT/QuNet, as well as from the JTF project NQuN (ID 60478) and from the EU H2020 Quantum Flagship projects QIA (820445) and QMiCS (820505). B.M. also acknowledges the support of H2020 project SPARTA, projects QuantMining POCI-01-0145-FEDER-031826, PREDICT PTDC/CCI-CIF/29877/2017 and QBigData PEst-OE/EEI/LA0008/2013, by FCT. The authors acknowledge the support from the project TheBlinQC supported by the EU H2020 QuantERA ERA-NET Cofund in Quantum Technologies and by FCT (QuantERA/0001/2017) and National Science Centre Poland No. 2017/25/Z/ST2/03027.

\begin{widetext}
\newpage
\begin{center}
\begin{table}[h]
\begin{tabular}{@{} p{4cm} p{0.5cm}  p{5cm}  p{0.5cm}  p{5cm} @{}}
\specialrule{.2em}{.1em}{.1em}  
\\
%\hline
\backslashbox{\large{\emph{Property}}}{\large{\emph{Top. effect}}} & \phantom{abc} &\large{\emph{Work quantization}} & \phantom{abc} & \large{\emph{Thouless pumping}}  
\\ \specialrule{0.1em}{0.1em}{0.1em} 
\\
Parameter space topologically  a torus $T^2$ & \phantom{abc} & Space-time &\phantom{abc} & Bloch momenta and time  \\ %\hline
\\
\\
Gapped Hamiltonian &\phantom{abc} & $M(t,x)=\sum_{\mu=1}^{3}B^{\mu}(t,x)J_{\mu}$ &\phantom{abc} & $H(t,k)=\sum_{\mu=1}^{3}d^{\mu}(t,k)\sigma_{\mu}$ (in the simplest scenario of a two-band system) \\ %\hline
\\
Wave-functions (sections) &\phantom{abc} & Dressed states $\ket{\eta(t,x)}$ &\phantom{abc} & Bloch states $\ket{u(t,k)}$ \\ %\hline
\\
Gauge field &\phantom{abc} & $-iA(t,x)/\hbar=\bra{\eta (t,x)}d\ket{\eta (t,x)}$ acts as an effective external field &\phantom{abc} & $A(t,k)=\bra{u(t,k)}d\ket{u(t,k)}$ manifests through coupling to an external field\\ %\hline
\\
Field strength & \phantom{abc} & $F(t,x)=dA$ &\phantom{abc} & $F(t,k)=dA$ \\ %\hline
\\
1st Chern number &\phantom{abc} & Average work performed by $E(t,x)$ on the unit cell of space-time lattice & \phantom{abc} & Shift of the center of mass of a system $x_{\text{cm}}=\langle x\rangle $ in one period of driving\\
\\
\specialrule{.2em}{.1em}{.1em} 
%\hline
\end{tabular}
\caption{Comparison between charge pumping and quantization of work. The $\sigma_{\mu}$'s denote the usual Pauli matrices.}
\label{tab: comparison}
\end{table}
\end{center}
\newpage
\end{widetext}

%%%%
\begin{widetext}

\section{Supplemental Material}
In this Supplemental Material, we first consider the topological properties of a general system described by a Hamiltonian linear in the Pauli matrices, which satisfy the $\mathfrak{su}(2)$-Lie algebra relations. Then, we discuss a generalization to an arbitrary representation of $\mbox{SU}(2)$ group and, finally, we show that the presented results immediately apply to the system considered in the Letter.
\subsection{Chern number and Dirac monopoles}
% and irreps of $\mbox{SU}(2)$}
Consider the two-level Hamiltonian
\begin{align}
H(x)=x^{\mu}\sigma_{\mu}, \text{ with } \delta_{\mu\nu}x^{\mu}x^{\nu}=1, \text{ i.e. } x\in S^2,
\label{eq: spin 1/2}
\end{align}
where $\{\sigma_{\mu}\}_{\mu=1}^{3}$ are the Pauli matrices and we have adopted the Einstein summation convention. The Pauli matrices satisfy the $\mathfrak{su}(2)$-Lie algebra relations 
\begin{align}
[\sigma_\mu,\sigma_{\nu}]=2i\varepsilon_{\mu\nu}^{\ \ \lambda}\sigma_{\lambda},
\label{eq: su(2)}
\end{align}
together with the Clifford algebra relations
\begin{align}
\sigma_{\mu}\sigma_{\nu}+\sigma_{\nu}\sigma_{\mu}=2\delta_{\mu\nu} I,
\label{eq: Clifford}
\end{align}
where $I$ denotes the $2\times 2$ identity matrix. The relations of Eq.~\eqref{eq: Clifford} imply that $\big(H(x)\big)^2=I$ and, thus, the eigenvalues of $H(x)$ are $\pm 1$. We can then consider the eigenspaces
\begin{align}
L_{x}=\{v\in\mathbb{C}^2: H(x) v=v\} \text{ and } L_{x}^{\perp}=\{v\in\mathbb{C}^2: H(x)v=-v\}, \text{ with } x\in S^2.
\end{align}
For each $x\in S^2$, there exists $U\in \mbox{SU}(2)$, such that
\begin{align}
H(x)=U\sigma_3 U^{-1}.
\label{eq: diag}
\end{align}
The first column of the matrix $U$ is just a choice of an element $v$ of $L_x$, with $||v||=1$, while the second column is $-i\sigma_2 \overline{v}$. This second choice ensures that $\det U(x)=1$. We can write $U$ in the form $U=[v ,\ -i\sigma_2\overline{v}]$. Another choice of $U$ is readily obtained by taking $v\to e^{i\alpha} v$, and this corresponding to taking $U(x)\to U\exp(i\alpha\sigma_3)$, which preserves Eq.~\eqref{eq: diag}. In fact, if we introduce the stereographic projection complex coordinate, with respect to the south pole of the sphere $x_0=(0,0,-1)\in S^2$,
\begin{align}
z=\frac{x^1+ix^2}{1+x^3},\ x\neq x_0,
\end{align}
we can take
\begin{align}
v(x)=\frac{1}{(1+|z|^2)^{1/2}}\left[\begin{array}{cc}
1 \\
z
\end{array}
\right],\ \text{ for } x\neq x_0,
\end{align}
as a smooth choice of $v(x)\in L_x$, with $||v(x)||=1$, for every $x\ne x_0$. One can introduce a complex coordinate $w=1/z$, corresponding to stereographic projection with respect to the north pole $x=-x_0$. And then, whenever $x\neq \pm x_0$
\begin{align}
v(x)=\frac{1}{(1+|z|^2)^{1/2}}\left[\begin{array}{cc}
1 \\
z
\end{array}
\right]=\frac{z}{|z|} \times  \frac{1}{(1+|w|^2)^{1/2}}\left[\begin{array}{cc}
w \\
1
\end{array}
\right]\equiv \frac{z}{|z|}\times v'(x),
\end{align}
and now $v'(x)$ is a choice valid for $x\neq -x_0$. The difference between the two choices is, whenever both are defined, i.e., $x\neq \pm x_0$, the gauge transformation $g(x)=z/|z| \in \mbox{U}(1)$.

It is impossible to find a global smooth choice of $v(x)$, which means that the line bundle over $S^2$,
\begin{align}
L=\coprod_{x\in S^2} L_x=\{(x,v): x\in S^2 \text{ and } v \in L_x\},
\end{align}
with fiber $L_x$ at $x\in S^2$, is not isomorphic to the trivial bundle $S^2\times \mathbb{C}$~\cite{mor:01}. This obstruction, topological in nature, is encoded in the gauge transformation $g(x)=z/|z|$. It is defined on $S^2-\{x_0,-x_0\}$, which is of the same homotopy type as the equator of $S^2$, topologically a circle $S^1=\{z\in \mathbb{C}:|z|=1\}$. Then, the transition map, seen as map from $S^1$ to $\mbox{U}(1)\cong S^1$, is nothing but the identity map, whose winding number is $1$. The Chern number of the line bundle $L$ is nothing but $-1$, i.e., minus the winding number of this transition map. The Chern number, being an integral of a characteristic class of $L$ (see Refs.~\cite{nak:03, mor:01}), measures the obstruction of $L$ being a trivial bundle. We can compute it by integrating the Berry curvature on the whole sphere $S^2$. Since we have two gauges $v(x)$ and $v'(x)$ related by a gauge transformation, we will have the corresponding local Berry gauge fields:
\begin{align}
A(x)=\bra{ v(x)}d \ket{v(x)}=\frac{1}{2}\frac{\bar{z}dz-zd\bar{z}}{(1+|z|^2)}, \text{ and } A'(x)=\bra{ v'(x)}d \ket{v' (x)}=\frac{1}{2}\frac{\bar{w}dw-wd\bar{w}}{(1+|w|^2)}.
\end{align}
On the overlap, they are related by
\begin{align*}
A(x)-A'(x)=d\log g(x).
\end{align*}
By taking the sphere to be the union of the northern and southern hemispheres, we see that the Chern number of $L$, is equal to, by Stokes' theorem,
\begin{align}
\int_{S^2} \frac{iF(x)}{2\pi}= \frac{i}{2\pi}\int_{S^1} (A(x)-A'(x))=\frac{i}{2\pi}\int_{S^1} \frac{dz}{z}=-1.
\end{align}
If we use coordinates of the ambient space $\mathbb{R}^3$ on which $S^2$ is embedded in, 
\begin{align}
F(x)=\frac{i}{4}\varepsilon_{\mu\nu\lambda}x^{\mu}dx^{\nu}\wedge dx^{\lambda},
\label{eq: Berry curvature}
\end{align}
which can be extended to a $\mbox{U}(1)$-field strength over $\mathbb{R}^3-\{0\}$:
\begin{align}
F(x)=\frac{d\bar{z}\wedge dz}{(1+|z|^2)^2}=\frac{i}{4}\varepsilon_{\mu\nu\lambda}\frac{x^{\mu}}{|x|^3}dx^{\nu}\wedge dx^{\lambda}.
\end{align}
Since this is $i/(2 R^2)$ times the area element of a sphere of radius $R$, we see that this corresponds to the field strength of a Dirac monopole of topological charge $-1$ sitting at the origin of $\mathbb{R}^3$, with the magnetic field given by
\begin{align}
B(x)=-\frac{1}{2}\frac{x}{|x|^3}.
\end{align}
and hence the flux over a sphere of radius $R$, $S_R=\{x\in \mathbb{R}^3:|x|=R\}$ with unit normal $n=x/|x|$, is
\begin{align}
\int_{S_R} \frac{iF(x)}{2\pi}=\frac{1}{2\pi}\int_{S_R} B(x)\cdot n(x) \ R^2 d\Omega=-\int_{S_R} \frac{d\Omega}{4\pi}=-1,
\end{align}
where $d\Omega$ denotes the infinitesimal solid angle.

\subsection{Irreducible representations of $\mbox{SU}(2)$}
Because of the $\mathfrak{su}(2)$-Lie algebra relations of Eq.~\eqref{eq: su(2)}, the construction presented in the previous section has an immediate generalization to an arbitrary representation of $\mbox{SU}(2)$, \emph{cf.}~\cite{dar:jam:12}. Notice that the diagonal subgroup $\mbox{U}(1)$ generated by $\sigma_3$, has the effect of gauge transformations. Indeed, given a local unitary gauge specified by $v(x)$, $U(x)=[v(x), -i\sigma_2 \overline{v}(x)]$, we have seen that multiplication by $\exp(i\alpha(x)\sigma_3)$ on the right is equivalent to multiplication of $v(x)$ by the gauge transformation $e^{i\alpha(x)}$. In general, given the $2J+1$-dimensional irreducible representation of $\mbox{SU}(2)$ of spin $J$, with generators $\{J_{\mu}\}_{\mu=1}^{3}$, we have the replacement
\begin{align}
\sigma_3\to 2 J_{3}=2 \ \mbox{diag}(J,J-1,....,-(J-1),-J).
\end{align}
This means that, if we write the Hamiltonian
\begin{align}
H(x)=x^{\mu}J_{\mu},
\end{align}
it will split the Hilbert space $\mathbb{C}^{2J+1}$ into $2J+1$ orthogonal sectors, corresponding to the eigenspaces for the different eigenvalues $\{-J,...,J\}$, of $H(x)$. In fact, the local gauge $U(x)=[v(x), -i\sigma_2 v(x)]$ in the spin-$\frac{1}{2}$ irreducible representation induces a local gauge in the spin-$J$ representation which is the image of $U(x)$ under the representation map ($\rho:\mbox{SU}(2)\to \mbox{U}(2J+1)$). In the sector specified by $J_3=m$ (which is the diagonal form of $H(x)$), a gauge transformation $U(x)\to U(x)\exp(2i\alpha(x) J_{3})$ will induce a gauge transformation
\begin{align}
v(x)\to v(x) \exp(2i\alpha(x) m).
\end{align}
From this, we can read off the induced Chern numbers: simply $-2m$ for the sector with $J_3=m$, $m\in\{-J,...,J\}$. For example, for spin $J=1$, we have three sectors, with Chern numbers $-2$, $0$ and $+2$, respectively.

In two-dimensional translation invariant symmetry protected topological phases of free fermions, we often encounter single particle Hamiltonians of the form of Eq.~\eqref{eq: spin 1/2}, depending on the quasi-momenta $\bf{k}$ in the Brillouin zone $\text{B.Z.}$ which topologically is a $2$-torus, $\text{B.Z.}\cong T^2$:
\begin{align*}
h(\bf{k})=x^{\mu}(\bf{k})\sigma_{\mu},\  \bf{k}\in T^2.
\end{align*}
The presence of gap means that for each $\bf{k}$, the vector $x(\bf{k})$ is non-vanishing. One can then smoothly deform the previous Hamiltonian, by an operation known as spectrum flattening, in such a way that $x(\bf{k})\in S^2$. This means that we have a map $x: T^2\to S^2$. The eigenspace of positive energy for momentum $\bf{k}$ is simply $L_{x(\bf{k})}$, for every $\bf{k}\in T^2$. This means that the eigenspaces of $h(\bf{k})$ are completely determined by those of $H(x)$ and the map $x$: $h(\bf{k})=H(x(\bf{k}))$. All the geometric structures can then be ``pulled back'' using $x$. In particular, the Berry curvature for the positive energy band has the form, see Eq.~\eqref{eq: Berry curvature}
\begin{align*}
F(\bf{k})=\frac{i}{2}\varepsilon_{\mu\nu\lambda}x^{\mu}(\bf{k})\frac{\partial x^{\nu}}{\partial k_x}(\bf{k})\frac{\partial x^{\lambda}}{\partial k_y}(\bf{k})dk_x\wedge dk_y=\frac{i}{2} x(\bf{k})\cdot \Big(\frac{\partial x}{\partial k_x}(\bf{k})\times \frac{\partial x}{\partial k_y}(\bf{k}) \Big)dk_x\wedge dk_y,
\end{align*}
yielding the formula for the Chern number
\begin{align*}
-\frac{1}{4\pi}\int_{B.Z.\cong T^2} x(\bf{k})\cdot \Big(\frac{\partial x}{\partial k_x}(\bf{k})\times \frac{\partial x}{\partial k_y}(\bf{k}) \Big)dk_x\wedge dk_y,
\end{align*}
which is minus the winding number of the map $x:T^2\to S^2$. From the previous arguments of representation theory, if we replace the spin-$1/2$ representation by spin-$J$, then, on the sector with $J_3=m$ of the irreducible representation of spin-$J$ of $\mbox{SU}(2)$, we will have a Chern number,
\begin{align}
-\frac{2m}{4\pi}\int_{B.Z.\cong T^2} x(\bf{k})\cdot \Big(\frac{\partial x}{\partial k_x}(\bf{k})\times \frac{\partial x}{\partial k_y}(\bf{k}) \Big)dk_x\wedge dk_y,
\end{align}
i.e., $-2m$ times the winding number of the map $x:T^2\to S^2$. 

\subsection{The system considered in the Letter}
% and irreps of $\mbox{SU}(2)$}
In the main text, the field $\bf{B}(t,x)$, for $\gamma/\nu \notin\{0,-1,1\}$, defines a map from a torus (defined by its periodicity in  time and space: $t\sim t+2\pi/\widetilde{\omega} $ and $x\sim x+2\pi/k$) to the sphere. Moreover, the  $J_{\mu}$'s appearing in the expression of the Hamiltonian $M(t,x)=\sum_{\mu}B^{\mu}(t,x)J_{\mu}$ define an irreducible representation of $\mbox{SU}(2)$ with spin $1$, hence, we will have three energy sectors corresponding to $J_3\in \{+1,0,-1\}$ with Chern numbers equal to $-2W$, $0$ and $+2W$, where
\begin{align}
W=\frac{1}{4\pi}\int_{0}^{2\pi/k}\int_{0}^{2\pi/\widetilde{\omega}} \frac{\bf{B}(t,x)\cdot \Big(\frac{\partial \bf{B}}{\partial t}(t,x)\times \frac{\partial \bf{B}}{\partial x}(t,x) \Big)}{|\bf{B}(t,x)|^3}dt \ dx
\label{winding number}
\end{align}
is the winding number of the induced map. Using Mathematica, we computed the winding number through the previous integral, with $\alpha/\nu=1$, obtaining
\begin{align}
W=\begin{cases}
0,\ \text{for } |\gamma/\nu|>1,\\
-2,\ \text{for } 0<\gamma/\nu<1,\\
+2,\ \text{for } -1<\gamma/\nu<0.
\end{cases}.
\end{align}
Hence, the positive energy band will have the Chern numbers claimed in the main text. The average, in one period of time, of the work performed by the synthetic field $E(t,x)$, to transport a particle by one space lattice spacing, is proportional to the winding number given by Eq.~\eqref{winding number}. Thus, the quantization of the average work presented in the Letter is a consequence of the fact that the corresponding Chern number is always integral: namely, in our case, an integer times the winding number of a map from the torus to the sphere.
\end{widetext}
%%%%%
%\bibliographystyle{unsrt}
\bibliography{bib}
\end{document}